\documentclass[a4paper,twocolumn,superscriptaddress,10pt,unpublished]{quantumarticle} 
\pdfoutput=1
\usepackage[utf8]{inputenc}
\usepackage[english]{babel}
\usepackage[T1]{fontenc}
\usepackage[numbers,sort&compress]{natbib}
\usepackage{amsmath}
\usepackage{amsthm}
\usepackage{amsfonts}
\usepackage{amssymb}
\usepackage{physics}
\usepackage{qcircuit}
\usepackage[ruled,vlined,linesnumbered]{algorithm2e}
\usepackage{booktabs} 
\usepackage{listings}
\usepackage{tikz} 
\usepackage{hyperref}

\numberwithin{equation}{section}

\lstdefinelanguage{Rust}{%
 sensitive%
,morecomment=[l]{//}%
,morecomment=[s]{/*}{*/}%
,moredelim=[s][{\itshape\color[rgb]{0,0,0.75}}]{\#[}{]}%
,morestring=[b]{"}%
,alsodigit={}%
,alsoother={}%
,alsoletter={!}%
, morekeywords={break, continue, else, for, if, in, loop, match, return, while} % control flow keywords
, morekeywords={as, const, let, move, mut, ref, static} % in the context of variables
, morekeywords={enum, fn, impl, Self, self, struct, trait, type, use, where} % in the context of declarations
, morekeywords={crate, extern, mod, pub, super} % in the context of modularisation
, morekeywords={unsafe} % markers
, morekeywords={abstract, alignof, become, box, do, final, macro, offsetof, override, priv, proc, pure, sizeof, typeof, unsized, virtual, yield} % reserved identifiers
, morekeywords=[2]{Add, AddAssign, Any, AsciiExt, AsInner, AsInnerMut, AsMut, AsRawFd, AsRawHandle, AsRawSocket, AsRef, Binary, BitAnd, BitAndAssign, Bitor, BitOr, BitOrAssign, BitXor, BitXorAssign, Borrow, BorrowMut, Boxed, BoxPlace, BufRead, BuildHasher, CastInto, CharExt, Clone, CoerceUnsized, CommandExt, Copy, Debug, DecodableFloat, Default, Deref, DerefMut, DirBuilderExt, DirEntryExt, Display, Div, DivAssign, DoubleEndedIterator, DoubleEndedSearcher, Drop, EnvKey, Eq, Error, ExactSizeIterator, ExitStatusExt, Extend, FileExt, FileTypeExt, Float, Fn, FnBox, FnMut, FnOnce, Freeze, From, FromInner, FromIterator, FromRawFd, FromRawHandle, FromRawSocket, FromStr, FullOps, FusedIterator, Generator, Hash, Hasher, Index, IndexMut, InPlace, Int, Into, IntoCow, IntoInner, IntoIterator, IntoRawFd, IntoRawHandle, IntoRawSocket, IsMinusOne, IsZero, Iterator, JoinHandleExt, LargeInt, LowerExp, LowerHex, MetadataExt, Mul, MulAssign, Neg, Not, Octal, OpenOptionsExt, Ord, OsStrExt, OsStringExt, Packet, PartialEq, PartialOrd, Pattern, PermissionsExt, Place, Placer, Pointer, Product, Put, RangeArgument, RawFloat, Read, Rem, RemAssign, Seek, Shl, ShlAssign, Shr, ShrAssign, Sized, SliceConcatExt, SliceExt, SliceIndex, Stats, Step, StrExt, Sub, SubAssign, Sum, Sync, TDynBenchFn, Terminal, Termination, ToOwned, ToSocketAddrs, ToString, Try, TryFrom, TryInto, UnicodeStr, Unsize, UpperExp, UpperHex, WideInt, Write}
, morekeywords=[2]{Send} % additional traits
, morekeywords=[3]{bool, char, f32, f64, i8, i16, i32, i64, isize, str, u8, u16, u32, u64, unit, usize, i128, u128} % primitive types
, morekeywords=[4]{Err, false, None, Ok, Some, true} % prelude value constructors
, morekeywords=[3]{AccessError, Adddf3, AddI128, AddoI128, AddoU128, ADDRESS, ADDRESS64, addrinfo, ADDRINFOA, AddrParseError, Addsf3, AddU128, advice, aiocb, Alignment, AllocErr, AnonPipe, Answer, Arc, Args, ArgsInnerDebug, ArgsOs, Argument, Arguments, ArgumentV1, Ashldi3, Ashlti3, Ashrdi3, Ashrti3, AssertParamIsClone, AssertParamIsCopy, AssertParamIsEq, AssertUnwindSafe, AtomicBool, AtomicPtr, Attr, auxtype, auxv, BackPlace, BacktraceContext, Barrier, BarrierWaitResult, Bencher, BenchMode, BenchSamples, BinaryHeap, BinaryHeapPlace, blkcnt, blkcnt64, blksize, BOOL, boolean, BOOLEAN, BoolTrie, BorrowError, BorrowMutError, Bound, Box, bpf, BTreeMap, BTreeSet, Bucket, BucketState, Buf, BufReader, BufWriter, Builder, BuildHasherDefault, BY, BYTE, Bytes, c, CannotReallocInPlace, cc, Cell, Chain, CHAR, CharIndices, CharPredicateSearcher, Chars, CharSearcher, CharsError, CharSliceSearcher, CharTryFromError, Child, ChildPipes, ChildStderr, ChildStdin, ChildStdio, ChildStdout, Chunks, ChunksMut, ciovec, clock, clockid, Cloned, cmsgcred, cmsghdr, CodePoint, Color, ColorConfig, Command, CommandEnv, Component, Components, CONDITION, condvar, Condvar, CONSOLE, CONTEXT, Count, Cow, cpu, CRITICAL, CStr, CString, CStringArray, Cursor, Cycle, CycleIter, daddr, DebugList, DebugMap, DebugSet, DebugStruct, DebugTuple, Decimal, Decoded, DecodeUtf16, DecodeUtf16Error, DecodeUtf8, DefaultEnvKey, DefaultHasher, dev, device, Difference, Digit32, DIR, DirBuilder, dircookie, dirent, dirent64, DirEntry, Discriminant, DISPATCHER, Display, Divdf3, Divdi3, Divmoddi4, Divmodsi4, Divsf3, Divsi3, Divti3, dl, Dl, Dlmalloc, Dns, DnsAnswer, DnsQuery, dqblk, Drain, DrainFilter, Dtor, Duration, DwarfReader, DWORD, DWORDLONG, DynamicLibrary, Edge, EHAction, EHContext, Elf32, Elf64, Empty, EmptyBucket, EncodeUtf16, EncodeWide, Entry, EntryPlace, Enumerate, Env, epoll, errno, Error, ErrorKind, EscapeDebug, EscapeDefault, EscapeUnicode, event, Event, eventrwflags, eventtype, ExactChunks, ExactChunksMut, EXCEPTION, Excess, ExchangeHeapSingleton, exit, exitcode, ExitStatus, Failure, fd, fdflags, fdsflags, fdstat, ff, fflags, File, FILE, FileAttr, filedelta, FileDesc, FilePermissions, filesize, filestat, FILETIME, filetype, FileType, Filter, FilterMap, Fixdfdi, Fixdfsi, Fixdfti, Fixsfdi, Fixsfsi, Fixsfti, Fixunsdfdi, Fixunsdfsi, Fixunsdfti, Fixunssfdi, Fixunssfsi, Fixunssfti, Flag, FlatMap, Floatdidf, FLOATING, Floatsidf, Floatsisf, Floattidf, Floattisf, Floatundidf, Floatunsidf, Floatunsisf, Floatuntidf, Floatuntisf, flock, ForceResult, FormatSpec, Formatted, Formatter, Fp, FpCategory, fpos, fpos64, fpreg, fpregset, FPUControlWord, Frame, FromBytesWithNulError, FromUtf16Error, FromUtf8Error, FrontPlace, fsblkcnt, fsfilcnt, fsflags, fsid, fstore, fsword, FullBucket, FullBucketMut, FullDecoded, Fuse, GapThenFull, GeneratorState, gid, glob, glob64, GlobalDlmalloc, greg, group, GROUP, Guard, GUID, Handle, HANDLE, Handler, HashMap, HashSet, Heap, HINSTANCE, HMODULE, hostent, HRESULT, id, idtype, if, ifaddrs, IMAGEHLP, Immut, in, in6, Incoming, Infallible, Initializer, ino, ino64, inode, input, InsertResult, Inspect, Instant, int16, int32, int64, int8, integer, IntermediateBox, Internal, Intersection, intmax, IntoInnerError, IntoIter, IntoStringError, intptr, InvalidSequence, iovec, ip, IpAddr, ipc, Ipv4Addr, ipv6, Ipv6Addr, Ipv6MulticastScope, Iter, IterMut, itimerspec, itimerval, jail, JoinHandle, JoinPathsError, KDHELP64, kevent, kevent64, key, Key, Keys, KV, l4, LARGE, lastlog, launchpad, Layout, Lazy, lconv, Leaf, LeafOrInternal, Lines, LinesAny, LineWriter, linger, linkcount, LinkedList, load, locale, LocalKey, LocalKeyState, Location, lock, LockResult, loff, LONG, lookup, lookupflags, LookupHost, LPBOOL, LPBY, LPBYTE, LPCSTR, LPCVOID, LPCWSTR, LPDWORD, LPFILETIME, LPHANDLE, LPOVERLAPPED, LPPROCESS, LPPROGRESS, LPSECURITY, LPSTARTUPINFO, LPSTR, LPVOID, LPWCH, LPWIN32, LPWSADATA, LPWSAPROTOCOL, LPWSTR, Lshrdi3, Lshrti3, lwpid, M128A, mach, major, Map, mcontext, Metadata, Metric, MetricMap, mflags, minor, mmsghdr, Moddi3, mode, Modsi3, Modti3, MonitorMsg, MOUNT, mprot, mq, mqd, msflags, msghdr, msginfo, msglen, msgqnum, msqid, Muldf3, Mulodi4, Mulosi4, Muloti4, Mulsf3, Multi3, Mut, Mutex, MutexGuard, MyCollection, n16, NamePadding, NativeLibBoilerplate, nfds, nl, nlink, NodeRef, NoneError, NonNull, NonZero, nthreads, NulError, OccupiedEntry, off, off64, oflags, Once, OnceState, OpenOptions, Option, Options, OptRes, Ordering, OsStr, OsString, Output, OVERLAPPED, Owned, Packet, PanicInfo, Param, ParseBoolError, ParseCharError, ParseError, ParseFloatError, ParseIntError, ParseResult, Part, passwd, Path, PathBuf, PCONDITION, PCONSOLE, Peekable, PeekMut, Permissions, PhantomData, pid, Pipes, PlaceBack, PlaceFront, PLARGE, PoisonError, pollfd, PopResult, port, Position, Powidf2, Powisf2, Prefix, PrefixComponent, PrintFormat, proc, Process, PROCESS, processentry, protoent, PSRWLOCK, pthread, ptr, ptrdiff, PVECTORED, Queue, radvisory, RandomState, Range, RangeFrom, RangeFull, RangeInclusive, RangeMut, RangeTo, RangeToInclusive, RawBucket, RawFd, RawHandle, RawPthread, RawSocket, RawTable, RawVec, Rc, ReadDir, Receiver, recv, RecvError, RecvTimeoutError, ReentrantMutex, ReentrantMutexGuard, Ref, RefCell, RefMut, REPARSE, Repeat, Result, Rev, Reverse, riflags, rights, rlim, rlim64, rlimit, rlimit64, roflags, Root, RSplit, RSplitMut, RSplitN, RSplitNMut, RUNTIME, rusage, RwLock, RWLock, RwLockReadGuard, RwLockWriteGuard, sa, SafeHash, Scan, sched, scope, sdflags, SearchResult, SearchStep, SECURITY, SeekFrom, segment, Select, SelectionResult, sem, sembuf, send, Sender, SendError, servent, sf, Shared, shmatt, shmid, ShortReader, ShouldPanic, Shutdown, siflags, sigaction, SigAction, sigevent, sighandler, siginfo, Sign, signal, signalfd, SignalToken, sigset, sigval, Sink, SipHasher, SipHasher13, SipHasher24, size, SIZE, Skip, SkipWhile, Slice, SmallBoolTrie, sockaddr, SOCKADDR, sockcred, Socket, SOCKET, SocketAddr, SocketAddrV4, SocketAddrV6, socklen, speed, Splice, Split, SplitMut, SplitN, SplitNMut, SplitPaths, SplitWhitespace, spwd, SRWLOCK, ssize, stack, STACKFRAME64, StartResult, STARTUPINFO, stat, Stat, stat64, statfs, statfs64, StaticKey, statvfs, StatVfs, statvfs64, Stderr, StderrLock, StderrTerminal, Stdin, StdinLock, Stdio, StdioPipes, Stdout, StdoutLock, StdoutTerminal, StepBy, String, StripPrefixError, StrSearcher, subclockflags, Subdf3, SubI128, SuboI128, SuboU128, subrwflags, subscription, Subsf3, SubU128, Summary, suseconds, SYMBOL, SYMBOLIC, SymmetricDifference, SyncSender, sysinfo, System, SystemTime, SystemTimeError, Take, TakeWhile, tcb, tcflag, TcpListener, TcpStream, TempDir, TermInfo, TerminfoTerminal, termios, termios2, TestDesc, TestDescAndFn, TestEvent, TestFn, TestName, TestOpts, TestResult, Thread, threadattr, threadentry, ThreadId, tid, time, time64, timespec, TimeSpec, timestamp, timeval, timeval32, timezone, tm, tms, ToLowercase, ToUppercase, TraitObject, TryFromIntError, TryFromSliceError, TryIter, TryLockError, TryLockResult, TryRecvError, TrySendError, TypeId, U64x2, ucontext, ucred, Udivdi3, Udivmoddi4, Udivmodsi4, Udivmodti4, Udivsi3, Udivti3, UdpSocket, uid, UINT, uint16, uint32, uint64, uint8, uintmax, uintptr, ulflags, ULONG, ULONGLONG, Umoddi3, Umodsi3, Umodti3, UnicodeVersion, Union, Unique, UnixDatagram, UnixListener, UnixStream, Unpacked, UnsafeCell, UNWIND, UpgradeResult, useconds, user, userdata, USHORT, Utf16Encoder, Utf8Error, Utf8Lossy, Utf8LossyChunk, Utf8LossyChunksIter, utimbuf, utmp, utmpx, utsname, uuid, VacantEntry, Values, ValuesMut, VarError, Variables, Vars, VarsOs, Vec, VecDeque, vm, Void, WaitTimeoutResult, WaitToken, wchar, WCHAR, Weak, whence, WIN32, WinConsole, Windows, WindowsEnvKey, winsize, WORD, Wrapping, wrlen, WSADATA, WSAPROTOCOL, WSAPROTOCOLCHAIN, Wtf8, Wtf8Buf, Wtf8CodePoints, xsw, xucred, Zip, zx}
, morekeywords=[5]{assert!, assert_eq!, assert_ne!, cfg!, column!, compile_error!, concat!, concat_idents!, debug_assert!, debug_assert_eq!, debug_assert_ne!, env!, eprint!, eprintln!, file!, format!, format_args!, include!, include_bytes!, include_str!, line!, module_path!, option_env!, panic!, print!, println!, select!, stringify!, thread_local!, try!, unimplemented!, unreachable!, vec!, write!, writeln!} % prelude macros
}%

\definecolor{comment}{RGB}{0,128,0}    % dark green
\definecolor{string}{RGB}{255,0,0}     % red
\definecolor{keyword}{RGB}{0,0,255}    % blue
\definecolor{function}{RGB}{128,0,128} % purple
\definecolor{type}{RGB}{128,0,0}       % dark red

\lstdefinestyle{go}{
  commentstyle=\color{comment},
  stringstyle=\color{string},
  keywordstyle=[1]\color{keyword},
  keywordstyle=[2]\color{function},
  keywordstyle=[3]\color{type},
  basicstyle=\footnotesize\ttfamily,
  numbers=left,
  numberstyle=\tiny,
  numbersep=5pt,
  frame=lines,
  breaklines=true,
  prebreak=\raisebox{0ex}[0ex][0ex]{\ensuremath{\hookleftarrow}},
  showstringspaces=false,
  tabsize=3,
}

\definecolor{backcolour}{rgb}{0.95,0.95,0.92}
\newcommand{\inlinecode}[2]{\colorbox{backcolour}{\lstinline[language=#1]$#2$}}
  
\begin{document}
\title{Simulating Quantum Computers Using OpenCL}
 
\date{\today}
\author{Adam Kelly}
\orcid{0000-0002-8490-8088}

\maketitle

\begin{abstract}
  Quantum computing is an emerging technology, promising a paradigm shift in computing, and allowing for speed ups in many different problems. However, quantum devices are still in their early stages, most with only a small number qubits. This places a reliance on simulation to develop quantum algorithms and to verify these devices. While there exists many algorithms for the simulation of quantum circuits,
  there is (at the time of writing) no tools which use OpenCL to parallelize
  this simulation, thereby taking advantage of devices such as GPUs while still remaining portable.

  In this paper, such a tool is described, including optimizations in areas such as gate application. This leads to a new approach that outperforms other popular state vector based simulators.
  An implementation of the proposed simulator is available at \url{https://qcgpu.github.io}.
\end{abstract}

\section{Introduction}

Quantum computing is a paradigm shift in computing.
These devices are thought to be the key to solving some types of problems,
such as factoring semi-prime integers \cite{shor_polynomial-time_1999},
search for elements in an unstructured database \cite{grover_fast_1996, zalka_grovers_1999},
simulation of quantum systems, optimization \cite{farhi_quantum_2014} and chemistry problems.

These problems are not feasible to solve using classical computers, but quantum computers may fix that.
Still, it is estimated that hundreds \cite{abrams_quantum_1999}
up to thousands \cite{beauregard_circuit_2002} of qubits (the quantum analogue to bits) will be needed.
Still, the way that quantum computers operate does not violate the Church-Turing principle \cite{deutsch_quantum_1985}.
This means that quantum computers can be, to some extent, simulated using classical computers.

There are some existing quantum computers, such as IBM's Q Experience \cite{noauthor_ibm_2018},
a semi-public cloud based quantum computer with up to 20 qubits.
While the number of qubits available at the moment is small, as it increases, many issues are being raised.
One of these issues is the ability to assess the correctness, performance and scalability of quantum algorithms.
It is this issue which simulators of quantum computers address. They allow the user to test quantum algorithms using a
limited number of qubits, and calculate measurements, state amplitudes and density matrices.

In this work, a simulator using OpenCL is described, a technology introduced in section \ref{opencl}.

\subsection{Existing Research}

The idea of using classical computers to simulate quantum computers and quantum mechanics is nothing new.
There exists a variety of software libraries that can be used to so, each with different purposes.
Some libraries such as QuTIP\cite{qutip} are aimed at solving a wide variety of quantum mechanical problems, whereas others are more specialized such as Quipper \cite{green_quipper_2013} for controlling quantum computers and qHipster for simulating quantum computers using distributed computing techniques \cite{smelyanskiy_qhipster_2016}.
A comprehensive list of tools is available on Quantiki \cite{quantiki}.

While the area of simulation is well established, there are, to my knowledge,
no simulation tools that can take advantage of hardware acceleration.
It is well known that dedicated hardware can speed up certain types of computations. This is becoming increasingly more apparent in fields such as machine learning, gaming and cryptocurrency mining.

While this research mainly looks at state vector simulations, there are other ways of doing simulations.
These include using the Feynmann path integral formulation of quantum mechanics \cite{boixo_simulation_2017, chen_classical_2018}, using tensor networks \cite{markov_simulating_2008} and applying different simulations for circuits made up of certain types of quantum gates \cite{bravyi_simulation_2018}.
These techniques (while not covered in this work) will hopefully be included in the simulation software at a later date.

\subsection{OpenCL} \label{opencl}

OpenCL (Open Computing Language) is a general-purpose framework for heterogeneous parallel computing on cross-vendor hardware, such as CPUs, GPUs, DSP (digital signal processors) and FPGAs (field-programmable gate arrays).
It provides an abstraction for low-level hardware routing and a consistent memory and execution model for dealing with massively-parallel code execution. This allows the framework to scale from embedded systems to hardware from Nvidia, ATI, AMD, Intel and other manufacturers, all without having to rewrite the source code for various architectures.
A more detailed overview of OpenCL is given in \cite{tompson2012introduction}.

The main advantage of using OpenCL over a hardware specific framework is that of a portability first approach. OpenCL has the largest hardware coverage, and as a header only library, it requires no specific tools or other dependencies.
Aside from this, OpenCL is very well suited to tasks that can be expressed as a program working in parallel over simple data structures (such as arrays/vectors).
The disadvantages with OpenCL, however, come from this lack of a hardware-specific approach.
Using proprietary frameworks can sometimes be faster than using OpenCL,
and sometimes it can also be more straightforward to develop kernels for the devices.

\begin{figure}[t]
	\centering
	\includegraphics[width=0.4\textwidth,height=0.4\textheight,keepaspectratio]{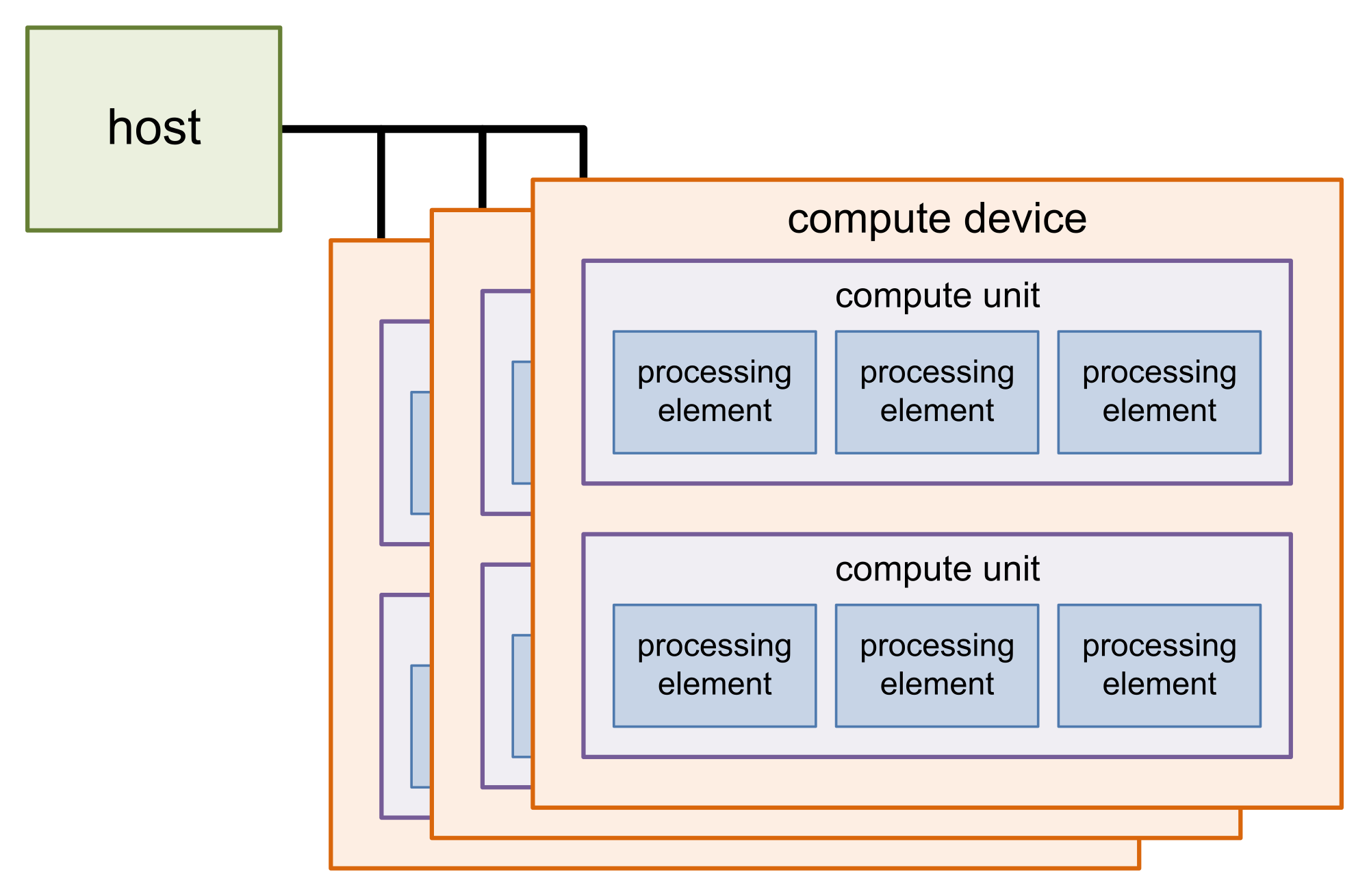}
	\caption{The OpenCL programming model/architecture}
	\label{fig:figure1}
\end{figure}

OpenCL is an open standard maintained by the non-profit Khronos Group. It views a computing system as a number of compute devices (such as CPUs or accelerators such as GPUs), attached to a host processor (a CPU). OpenCL executes functions on these devices called Kernels, and these kernels are written in a C-like language, OpenCL C.
A compute device is made up of several compute units which contain multiple processing elements.
It is the processing elements that execute kernels. This is shown in figure \ref{fig:figure1}.

At the host level, a compute device is selected. The OpenCL API then uses its platform later to submit work to the device and manage things like the work distribution and memory. The work is defined using kernels.
These kernels are written in OpenCL C, and execute in parallel over a predefined, $n$-dimensional computation domain. Each independent element of this execution is a work item.
These are equivalent to Nvidia CUDA threads. The groups of work items, work groups, are equivalent to CUDA thread blocks.

With this, a general pipeline for most GPGPU OpenCL applications can be described.
First, a CPU host defines an $n$-dimensional computation domain over some region of DRAM memory. Every index of this $n$-dimensional domain will be a work item, and each work item will execute the same given Kernel.

The host then defines a grouping of these into work groups. Each work item in the work-groups will execute concurrently within a compute unit and will share some local memory. These are placed on a work queue.

The hardware will then load DRAM into the global device RAM, and execute each work group on the work-queue.

On the device, the multiprocessor will execute the kernel using multiple threads at once. If there is more work groups than threads on the device, they will be serialized.

There are some limitations. The global work size must be a multiple of the work group size. This is to say the work group must fit evenly into the data structure.

Secondly, the number of elements in the $n$-dimensional vector must be less or equal the
\inlinecode{C}{CL_KERNEL_WORK_GROUP_SIZE} flag. This is important to the QCGPU library as it sets a hard limitation on the size of the state vector being stored on the GPU. \inlinecode{C}{CL_KERNEL_WORK_GROUP_SIZE} is a hardware flag, and OpenCL will return an error code if either of these conditions is violated.
This can be avoided by using an approach similar to the distributed memory techniques used in other simulations.
This feature is planned to be implemented soon.

\subsection{Quantum Computing}
\label{sec:quantum}

Before considering quantum computing, let's first start with classical computation.
A classical computer is the type of computer that you may have at home.
Laptops, Tablets, Phones and Smart TV's are all examples of a classical computer.

A quantum computer is different.
It takes advantage of principles of quantum mechanics such as superposition, entanglement and measurement to perform computation (see the following section).
Because of this, it can do computations that normal computers will never be able to do.

\subsubsection{Qubits and State}

In a classical computer, information is represented as a bit.
A bit is a binary system, and thus can be in one of two states,
0 or 1.
In a quantum computer, information is represented as a \emph{qubit}.
The qubit is the quantum analogue of a bit.
Using Dirac notation \cite{dirac1939new},
a qubit can be in the state $\ket{0}$ or $\ket{1}$, or (more importantly)
a \emph{superposition} (linear combination) of these states.
Mathematically, the state of a single qubit $\ket{\psi}$ is
\begin{equation}
	\ket{\psi} = \alpha \ket{0} + \beta \ket{1},
\end{equation}
\noindent
such that $\alpha, \beta \in \mathbb{C}$.
The coefficients also must follow a normalization condition of $\abs{\alpha}^2 + \abs{\beta}^2 = 1$.

In the above state, the complex numbers $\alpha$ and $\beta$ are known as \emph{amplitudes}.
The states $\ket{0}$ and $\ket{1}$ are known as basis states.
Importantly, given any state $\ket{\psi}$,
it is impossible to extract the amplitudes of any basis state.

Commonly used is the vector notation for states.
The basis states $\ket{0}$ and $\ket{1}$ are vectors that form an orthonormal basis for that qubits state space.
The standard representation (and the one followed throughout QCGPU and this paper) is

\begin{align}
	\ket{0} & = \begin{pmatrix}1 \\ 0\end{pmatrix}, &
	\ket{1} &= \begin{pmatrix}0 \\ 1\end{pmatrix}.
\end{align}

Following from this, the state $\ket{\psi}$ can be represented as a unit vector in the two-dimensional complex vector space,

\begin{equation}
	\ket{\psi} = \begin{pmatrix}\alpha \\ \beta\end{pmatrix}
\end{equation}

The concepts here generalize to quantum systems containing many qubits.
Since a single qubit has two distinct basis states,
an $n$ qubit system has $2^n$ distinct basis states.
In quantum computing, a multiple qubit system is known as a register.

To combine the states of two individual qubits, the Kronecker/tensor product must be used.
For example, to combine the states two qubits $\ket{\psi_1}$ and $\ket{\psi_2}$,
\begin{equation}
	\ket{\psi_1} \otimes \ket{\psi_2}
	= \begin{pmatrix}\alpha_1 \\ \beta_1 \end{pmatrix}  \otimes \begin{pmatrix}\alpha_2 \\ \beta_2 \end{pmatrix}
	% = \begin{pmatrix}\alpha_1 \cdot \begin{pmatrix}\alpha_2 \\ \beta_2 \end{pmatrix}  \\ \beta_1 \cdot \begin{pmatrix}\alpha_2 \\ \beta_2 \end{pmatrix} \end{pmatrix}
	= \begin{pmatrix}\alpha_1 \alpha_2 \\ \alpha_1 \beta_2  \\ \beta_1 \alpha_2 \\ \beta_1 \beta_2 \end{pmatrix}
\end{equation}

When basis vectors are combined, it is convention to say $\ket{1} \otimes \ket{0} = \ket{10}$ or $\ket{2}$ (as `10' is 2 in binary).

More generally, An $n$ qubit register is described by a unit vector $\ket{\phi}$ in the $2^n$ dimensional complex vector space,
\begin{equation}
	\ket{\phi} = \begin{pmatrix}\alpha_0 \\ \alpha_1 \\ \vdots \\ \alpha_{2^n - 1}\end{pmatrix}.
\end{equation}
\noindent
This is equivalent to a linear combination of the basis states
\begin{equation}
	\ket{\psi} = \sum_{j = 0}^{2^n - 1} \alpha_j \ket{j}
\end{equation}
\noindent
Where $\ket{j}$ is the $j$th basis vector, and $\sum \alpha_j = 1$.\\

There are some things note from this.
Consider the vector $\ket{\psi} = \frac{1}{\sqrt{2}}(\ket{00} + \ket{11})$.
It was stated before that individual qubits can be combined using the Kronecker/tensor product.
Yet, there is no solution for the vectors $\ket{a}$ and $\ket{b}$ to the equation $\ket{a} \otimes \ket{b} = \ket{\phi}$.
That is because $\ket{\psi}$ is \emph{entangled}, which means the state cannot be separated into individual qubit states.
This is important, as it is the entanglement that makes the simulation of quantum computers hard, as it means the number of
amplitudes that need to be stored grows exponentially rather then linearly.

\subsubsection{Manipulating the State}
\label{sec:gates}

In a classical computer, bits are manipulated using logic gates.
There is a quantum analogue to this too.

Just as the state of a system of qubits was defined using vectors,
the way they change can be described also.
The state of a qubit (or multiple qubits) is changed by quantum logic gates, or just gates.
When representing the state of qubits as vectors, quantum gates are represented using matrices.
These matrices must comply with certain rules in order to be valid quantum gates.

For a matrix to represent a quantum gate, it must be \emph{unitary}.
A matrix $U$ is unitary if it satisfies the property that it's conjugate transpose $U^\dagger$ is also its inverse,
thus $U^\dagger U = U U^\dagger = I$, where $I$ is the identity matrix.
In quantum computing, all gates have a corresponding unitary matrix, and all unitary matrices have a corresponding quantum gate.

Gates that act on a single qubit are represented by a $2 \times 2$ matrix.
More generally, an $n$ qubit gate is represented by a $2^n \times 2^n$ matrix.

A single qubit gate can be applied to a quantum register with an arbitrary number of qubits.
For a gate $U$ to act on the $j$th qubit in an $n$ qubit register, the full gate is formed by

\begin{equation}
	U = \underbrace{I \otimes I \otimes \dots }_{\text{$j - 1$ times}}\otimes U \otimes \underbrace{\dots \otimes I}_{\text{$n - j$ times}},
\end{equation}
\noindent
or more succinctly
\begin{equation}
	U_t = \bigotimes_{j = 1}^n \begin{cases}
		U & j = t     \\
		I & otherwise
	\end{cases}
\end{equation}

In matrix form, gates are applied to registers using matrix multiplication.
Multiple gates can be applied to a register. This is called a circuit. The gates being applied to a register can be detailed using a circuit diagram.

In a circuit diagram, each line across represents a qubit, and each of the blocks on the lines) represents gates or other operations such as measurement (see section \ref{sec:measure}).
An example circuit diagram for three qubits, applying the gate $U$ to the second qubit is shown below.

\vspace*{2\baselineskip}
\hfill
\Qcircuit @C=1em @R=1em {
\lstick{\ket{0}} & \qw & \qw & \rstick{\ket{0}} \qw \\
\lstick{\ket{0}} & \qw & \gate{U} & \rstick{U\ket{0}} \qw \\
\lstick{\ket{0}} & \qw & \qw & \rstick{\ket{0}} \qw
}
\hspace*{\fill}
\vspace*{2\baselineskip}

% There are a number of gates that are very common and are implemented in the software (among others).
% These are shown in table \ref{tab:gates}.

% \begin{table}[h]
% 	\centering
% 	\begin{tabular}{@{}lcl@{}} \toprule
% 		Gate & Circuit Diagram & Matrix  \\
% 		\toprule
% 		\textbf{Hadamard} & \Qcircuit @C=1em @R=.7em { & \gate{H} & \qw} & $H = \frac{1}{\sqrt{2}} \begin{bmatrix}
% 				1 & 1  \\
% 				1 & -1
% 			\end{bmatrix}$ \\
% 		\textbf{Pauli-X}  & \Qcircuit @C=1em @R=.7em { & \gate{X} & \qw} & $X = \begin{bmatrix}
% 				0 & 1 \\
% 				1 & 0
% 			\end{bmatrix}$                    \\
% 		\textbf{Pauli-Y}  & \Qcircuit @C=1em @R=.7em { & \gate{Y} & \qw} & $Y = \begin{bmatrix}
% 				0 & -i \\
% 				i & 0
% 			\end{bmatrix}$                    \\
% 		\textbf{Pauli-Z}  & \Qcircuit @C=1em @R=.7em { & \gate{Z} & \qw} & $Z = \begin{bmatrix}
% 				1 & 0  \\
% 				0 & -1
% 			\end{bmatrix}$                    \\
% 		\textbf{S}        & \Qcircuit @C=1em @R=.7em { & \gate{S} & \qw} & $S = \begin{bmatrix}
% 				1 & 0 \\
% 				0 & i
% 			\end{bmatrix}$                    \\
% 		\textbf{T}        & \Qcircuit @C=1em @R=.7em { & \gate{T} & \qw} & $T = \begin{bmatrix}
% 				1 & 0          \\
% 				0 & e^{i\pi/4}
% 			\end{bmatrix}$                    \\
% 		\textbf{CNOT}  & \Qcircuit @C=1em @R=.7em {
% 		& \ctrl{1} & \qw \\
% 		& \targ & \qw
% 		} & $CNOT = \begin{bmatrix}
% 				1 & 0 & 0 & 0 \\
% 				0 & 1 & 0 & 0 \\
% 				0 & 0 & 0 & 1 \\
% 				0 & 0 & 1 & 0
% 			\end{bmatrix}$  \\
% 		\bottomrule
% 	\end{tabular}
% 	\caption{Common quantum gates and their representations}
% 	\label{tab:gates}
% \end{table}

\subsubsection{Measurement}
\label{sec:measure}

It was stated before that given any state $\ket{\psi}$,
it is impossible to extract the amplitudes for each of the basis states.
Still, there is a way to get classical information, a bit, out of a qubit.

In the previous section, it was said that quantum states are altered by unitary transformations or matrices.
However, that only applies to a closed quantum system, that is, one that doesn't interact with external physical systems.
If you go to find out information about this quantum system, you are interacting with it.
This interaction causes the system to be no longer closed, and the system is no longer only altered by Unitary transformations.
This different type of interaction is called a measurement.

Quantum measurements are described by what is called a \emph{measurement operator}.
These are operators act on the vector space made up of the basis states of the quantum system being considered.
Measurement operators are a collection $\{ M_m \}$, where $m$ refers to the measurement outcome that may occur.

If a state of a quantum system (like a quantum register) is $\ket{\psi}$ immediately before a measurement,
then the probability of getting a result $m$ is given by

\begin{equation}\label{eq:probability}
	p(m) = \ev{M^\dagger_m M_m}{\psi},
\end{equation}
\noindent
and the state of the system after measurement, $\ket{\psi '}$ is

\begin{equation}
	\ket{\psi '} = \frac{M_m \ket{\psi}}{\sqrt{\ev{M^\dagger_m M_m}{\psi}}}.
\end{equation}

This description of measurement applies to an arbitrary quantum system, but now just qubits will be considered.
Qubits are almost always measured in the \emph{computational basis}. The measurement of a single qubit in the computational basis
has two measurement operators, $M_0 = \dyad{0}$, and $M_1 = \dyad{1}$. This means that there is two possible measurement outcomes, 0 and 1.

Now consider the state $\ket{\psi} = \alpha \ket{0} + \beta \ket{1}$.
Then, following from equation \ref{eq:probability}, the probability of obtaining a 0 when measuring is

\begin{equation}
	p(0) =  \ev{M^\dagger_0 M_0}{\psi} = \ev{M_0}{\psi} = \abs{\alpha}^2.
\end{equation}

In the same way, the probability of obtaining a 1 is $p(1) = \abs{\beta}^2$.
After the measurement, the two possible resulting states are:

\begin{align}
	\frac{M_0 \ket{\psi}}{\abs{\alpha}} = \frac{\alpha}{\abs{\alpha}}\ket{0} = \ket{0} \label{eq:phase} \\
	\frac{M_1 \ket{\psi}}{\abs{\beta}}  = \frac{\beta}{\abs{\beta}}\ket{1} = \ket{1} \label{eq:phase:2}
\end{align}
\noindent
Note that the coefficients in equation \ref{eq:phase} and \ref{eq:phase:2} and below are of the form $\frac{x}{\abs{x}}$.
This is equal to $\pm 1$. This can only be a global phase shift, and thus doesn't affect the measurement outcomes, and can be ignored. 

The principles shown here generalize to multiple qubits analogously, except there are $2^n$ possible measurement outcomes,
corresponding to the number of resulting basis states.

\section{Simulating Quantum Computers Using OpenCL}
\label{cha:implementation}

This section describes the simulation method used in the QCGPU library.
The focus will be on the OpenCL Implementations.

To be able to simulate a quantum computer, a simulation tool must have (at the bare minimum) a few things.
The first is the ability to represent the state of the quantum computer.
This is usually done by representing the state of the qubit register being considered,
and is discussed in section \ref{sec:state}.
Secondly, there needs to be a way to perform operations.
This is discussed in sections \ref{sec:repr_gates} and \ref{sec:gate_app}.
Lastly there needs to be a way to see the outcome of the operations.
This is usually done by implementing quantum measurement,
as discussed in section \ref{sec:measure_app}. However, it is sometimes useful to just see the unmeasured quantum state.
This is implemented in the simulator but is not discussed.

Throughout the software, the library `pyopencl' has been used to interact with OpenCL from python.
`numpy' is used throughout also.

\subsection{Representing State}
\label{sec:state}

As previously described, the state of an $n$ qubit register is characterized by a normalized vector in the $2^n$ dimensional complex vector space.
Because of this, such a state can be represented using $2^n$ complex numbers.
It is here that the main challenge with simulating quantum computers lies,
the exponential growth in the amount of complex numbers needed to describe a register.

In QCGPU, the state vector is stored as an array of $2^n$ complex floats.
These complex floats correspond to the components of the state vector.

When a new state is initialized with a given number of qubits, the initial state is $\ket{00\dots 0}$.
In terms of OpenCL, the array is stored on the device in global memory, with read and write permissions.

It should be noted the amount of memory needed to store an $n$ qubit state.
A complex float requires 64 bits, and the state is described by $2^n$ complex numbers, thus total amount of memory
needed to store the state vector is

\begin{equation}
	64 \cdot 2^n \text{bits}.
\end{equation}

To give a general idea, to simulate 5 qubits, 256 bytes are required. To simulate 10 qubits, 8.192 kilobytes are required.
To simulate 20 qubits, 8.389 megabytes are required. For 25 qubits, 268.4 megabytes are required and to simulate 32 qubits,
and for 30 qubits, 8.59 gigabytes of memory is required.

\subsection{Representing Gates}
\label{sec:repr_gates}

As the state of the qubits is represented as a vector, gates are represented as matrices.
This was looked at in section \ref{sec:quantum}.

It was stated before that to apply a single qubit gate $U$ to the $t$th qubit in an
$n$ qubit quantum register, the full matrix could be calculated by

\begin{equation}
	U_t = \bigotimes_{j = 1}^n \begin{cases}
		U & j = t     \\
		I & otherwise
	\end{cases}
\end{equation}

This presents a problem however, as it is very inefficient.
The first problem is that the size of such a matrix would be $2^n\times 2^n$.
That would take up a massive amount of memory, which is already a problem.
Secondly, this calculation relies on the Kronecker product, which for two matrices of size $n_1\times m_1$ and $n_2 \times m_2$,
has a running time of $O(n_1 n_2 m_1 m_2)$ using big-O notation. This would make the simulator extremely slow.

To avoid these issues, one has to use a different gate application algorithm to matrix multiplication (see the following section),
and represent gates in a different way.

In QCGPU, gates are stored as $2 \times 2$ matrices, and the only type of gates are single qubit gates.
From this, controlled gates (for multiple qubits) can be applied using any single qubit gate (again, see the following section).

This is possible due to a concept known as universality.
A set of gates is known as universal, if any possible operation on a quantum computer can be reduced to them.
An example of these sets is $\{ T, H, CNOT \}$.

The $T$ and $H$ gates are single qubit gates, thus can be represented in QCGPU,
and the $CNOT$ gate is just the controlled $X$ gate.
The $X$ gate is a single qubit gate, so it can be applied as a controlled gate using the software.
This means that the simulator can do any operation by just implementing single qubit gates and the ability to apply them as controlled gates.

For the implementation of the gates, it was chosen just to pass in each element of the 2x2 matrix into the OpenCL kernels.
This avoided complexity in the gate application methods. This can be seen in the following section.

In the library, single qubit gates are represented as a class.
This class allows the end user to just use either 2x2 arrays or 2x2 matrices from numpy to represent gates,
so as to not have to think about the internal representation.

\subsection{Improving the Gate Application Algorithm}
\label{sec:gate_app}

As gates are only represented as $2\times 2$ matrices, they can't be applied via matrix multiplication.
This means a different gate application algorithm must be used.

Algorithm \ref{alg:app} details this approach.
The structure of the algorithm is a for loop through have the number of amplitudes.
Note that the inside of the for loop is independent (not based on the rest of the computation).
This is what makes it suited to be parallelled. For the kernel source code, see appendix \ref{adx:gate}.

\begin{algorithm}[h]
	\KwIn{An $n$ qubit quantum state represented by a column vector $v = (v_1, \dots v_{2^n})^T$ and a single qubit gate $G$, represented by a $2\times 2$ matrix, acting on the $t$th qubit.}
	\BlankLine
	\For{$i \leftarrow 0$ \KwTo $2^{n - 1}$}
	{
		$a \leftarrow$ the $i$th integer who's $t$th bit is 0\;
		$b \leftarrow$ the $i$th integer who's $t$th bit is 1\;

		\tcp{The following must be simultaneously updated}
		$v_a \leftarrow v_a \cdot G_{0,0} + v_b \cdot G_{0, 1}$\;
		$v_b \leftarrow v_b \cdot G_{1,1} + v_a \cdot G_{1, 0}$\;
	}
	\caption{G\textsc{ate application Algorithm} ($v, G, t$)}
	\label{alg:app}
\end{algorithm}

To apply a single qubit gate as a controlled gate, algorithm \ref{alg:app} can be adapted.
If the control qubit is $c$th in the register, only apply the update to $v_a$ if the $c$th bit of $a$ is one,
and only update $v_b$ if the $c$th bit of $b$ is one. The corresponding kernel for this is shown in appendix \ref{adx:control}.

\subsection{Parallelizing the Measurement Algorithm}
\label{sec:measure_app}

The measurement process relies on knowing the probability of each output state.
The actual selection of an outcome based on these probabilities cannot be parallelizing,
however the calculation of the probabilities can.
For the source code, see appendix \ref{adx:measure}.

From this an outcome can be selected.
Because the probabilities can be calculated separately to the measurement,
it also allows multiple measurements to be made without having to apply all of the gates again.
While this isn't possible on a quantum computer,
it does mean that it is easier to prototype / simulate algorithms, the primary goal of the software library.

\section{Benchmarking}
\label{cha:benchmarking}

In order to see if using hardware acceleration to simulate quantum computers is faster then the conventional state vector approach,
it was necessary to benchmark the software against other commonly used tools.

It was decided to test against two different tools, ProjectQ \cite{steiger_projectq_2016} and the simulator in Qiskit \cite{noauthor_qiskit_nodate}.

\subsection{Designing The Experiments}

The goal of the benchmarking experiments was to test if there was a difference in speed between the different simulators.
The experiments were designed with this goal in mind.

\subsubsection{Avoiding Possible Errors}

The task of benchmarking software is not an easy one.
There are many different things which can impact the performance of software, all of which have to be taken into account when performing benchmarks.
Sometimes, the way that programming languages work,
different run-time optimizations can change the speed of some software.
This can be detrimental to the overall benchmarking results, and can be hard to diagnose.

Most of these issues boil down to independence.
The easiest way to avoid these issues is shuffling.
If you have a series of experiments to be run using the different tools,
the order in which each individual experiment is run should be random.
This is done in the benchmarking code in section \ref{sec:bench_method}.

\subsubsection{Reproducibility}

Reproducibility is very important in software benchmarking.
Different hardware and software configurations can make software change in performance.

To avoid this, all of the experiments were run using a virtual machine hosted by Amazon Web Services.
The machine was an EC2 P3.2xLarge instance with the following specifications:

\begin{table}[h]
	\centering
	\begin{tabular}{@{}lrrrrc@{}} \toprule
		                    & P3.2xLarge        \\
		\toprule
		\textbf{GPU}        & Nvidia Tesla V100 \\
		\textbf{GPU Memory} & 16GB              \\
		\textbf{vCPUs}      & 8                 \\
		\textbf{Memory}     & 61GB              \\
		\bottomrule
	\end{tabular}
	\caption{EC2 Instance Specifications}
	\label{tab:ec2}
\end{table}

\subsection{Benchmarking Method}
\label{sec:bench_method}

For the actual experiment that was being timed in the benchmarked, it was decided to use the quantum Fourier transform.
This is a transformation that can be built up using both single and controlled qubit gates.
The reasoning for using the quantum Fourier transform was that it is an integral part of many different quantum algorithms,
and thus would be a realistic task that the simulator would perform.

The benchmarking algorithm is detailed in algorithm \ref{alg:benchmarking}, and the benchmarking source code is given in the appendix.

\begin{algorithm}
	\KwIn{The number of qubits $n$ to test up to, and the number of samples for each number of qubits, $samples$}
	\KwOut{A list of the type of simulator, the number of qubits and the time taken to run the benchmark.}
	\BlankLine
	data $\leftarrow []$\;
	\For{$i \leftarrow 0$ \KwTo $n$}
	{
		\For{0 \KwTo $samples$}
		{
			type $\leftarrow$ a type of simulator to use, randomly chosen\;
			$t \leftarrow$ the time taken to run a quantum Fourier transform with $i$ qubits\;
			data $\leftarrow (\text{type}, i, t)$\;
		}
	}
	\Return{data}
	\caption{B\textsc{enchmarking algorithm} ($n$, $samples$)}
	\label{alg:benchmarking}
\end{algorithm}

\subsection{Results}

When running the benchmarks, it was found that after 24 qubits,
the IBM software was intermittent, occasionally throwing errors so it was decided to stop the benchmarks at the 24 qubit mark.

\begin{figure}[h]
	\centering
	\includegraphics[width=0.95\linewidth]{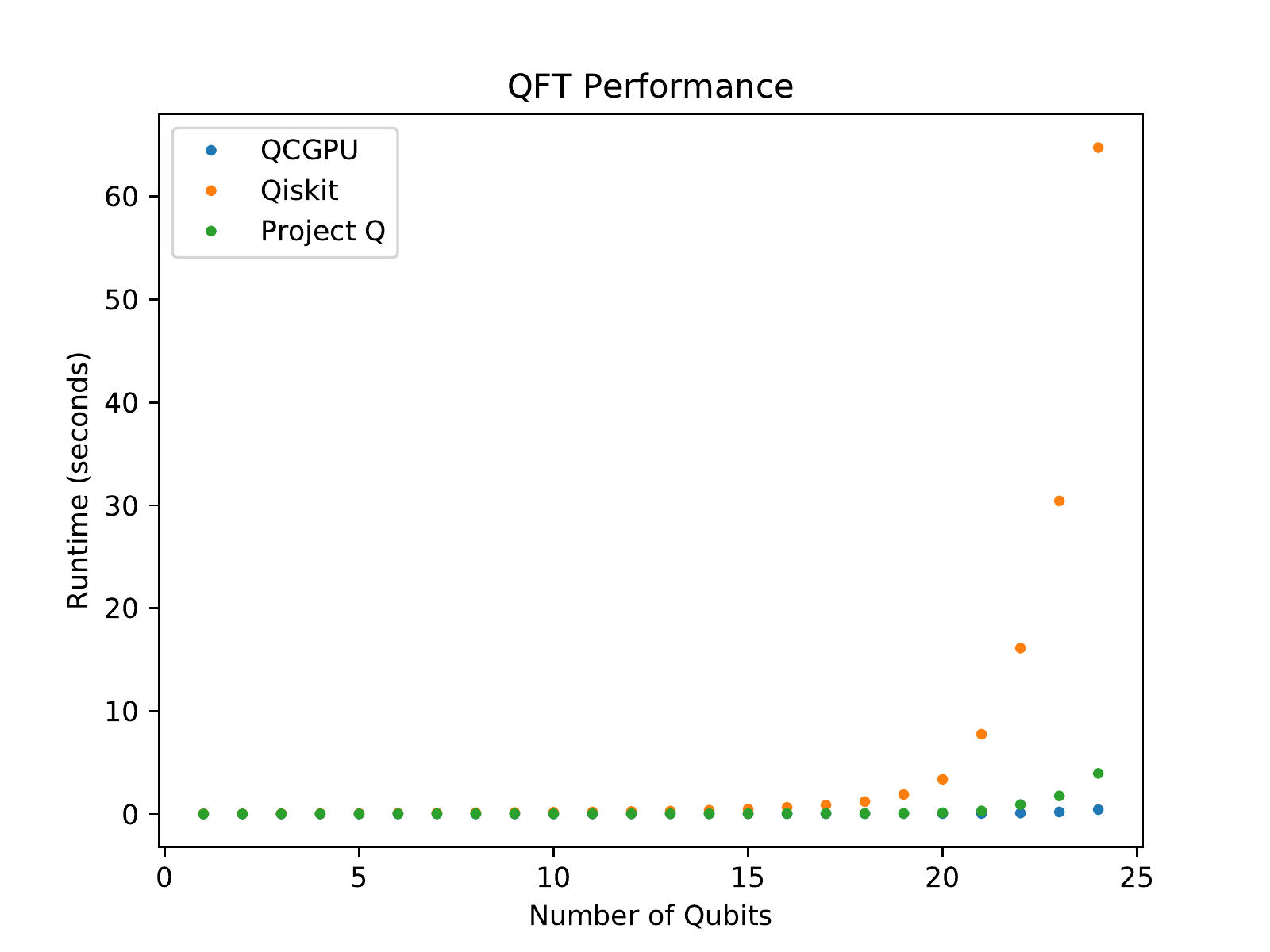}
	\caption{Benchmarking Data}
	\label{fig:bench}
\end{figure}

To see a graph of the mean running time for each simulator at between 1 and 24 qubits, see figure \ref{fig:bench}.

The biggest difference in time can be seen toward the end, where QCGPU is on average over 150 times faster than the Qiskit simulator
and 8 times faster then the ProjectQ simulator. This difference would only increase with larger circuits.

\subsubsection{A Statistical Analysis}

To prove the hypothesis of `using hardware acceleration provides a speed improvement over existing tools',
one needs to perform a statistical analysis.
The software was being compared against two tools, thus the analysis will be repeated twice, analogously.

The significance test for the populations was chosen based on the properties of the data set.

The dataset showed (against both Qiskit and ProjectQ) a significance in the homogeneity of variances.
This was determined using a Levene test, which gave p-values of 0.00194 and 0.006 respectively.

Using a Shapiro-Wilk test, it was found that the samples came from a normally distributed population, with p values of
0.0144 for QCGPU, 0.0333 for ProjectQ and 0.08 for Qiskit.

Because of these two properties, it was decided to use Welch's t-test to determine the p-value of the null hypothesis.
The resulting p-values were 0.0003396 when testing against qiskit, and 0.003189 when testing against projectq, thus the null hypothesis can be rejected.

\section{Conclusions}

The previous chapters have explored the implementation
of a library for the simulation of quantum computers, using hardware acceleration through the OpenCL framework.
Although time-consuming, the simulation of quantum computers is a necessary part of developing and testing new quantum algorithms.

Through the development of the library, it has been shown how hardware acceleration with devices such as GPUs can help speed up the simulation of quantum computers.
With the various optimizations done also, there has been shown to be a speedup, even on relatively low powered hardware, compared to existing libraries for a similar purpose.

\subsection{Applications of this Work}

The software developed during this research, QCGPU, has a number of very useful applications.

\begin{figure}
	\centering
	\includegraphics[width=0.8\linewidth]{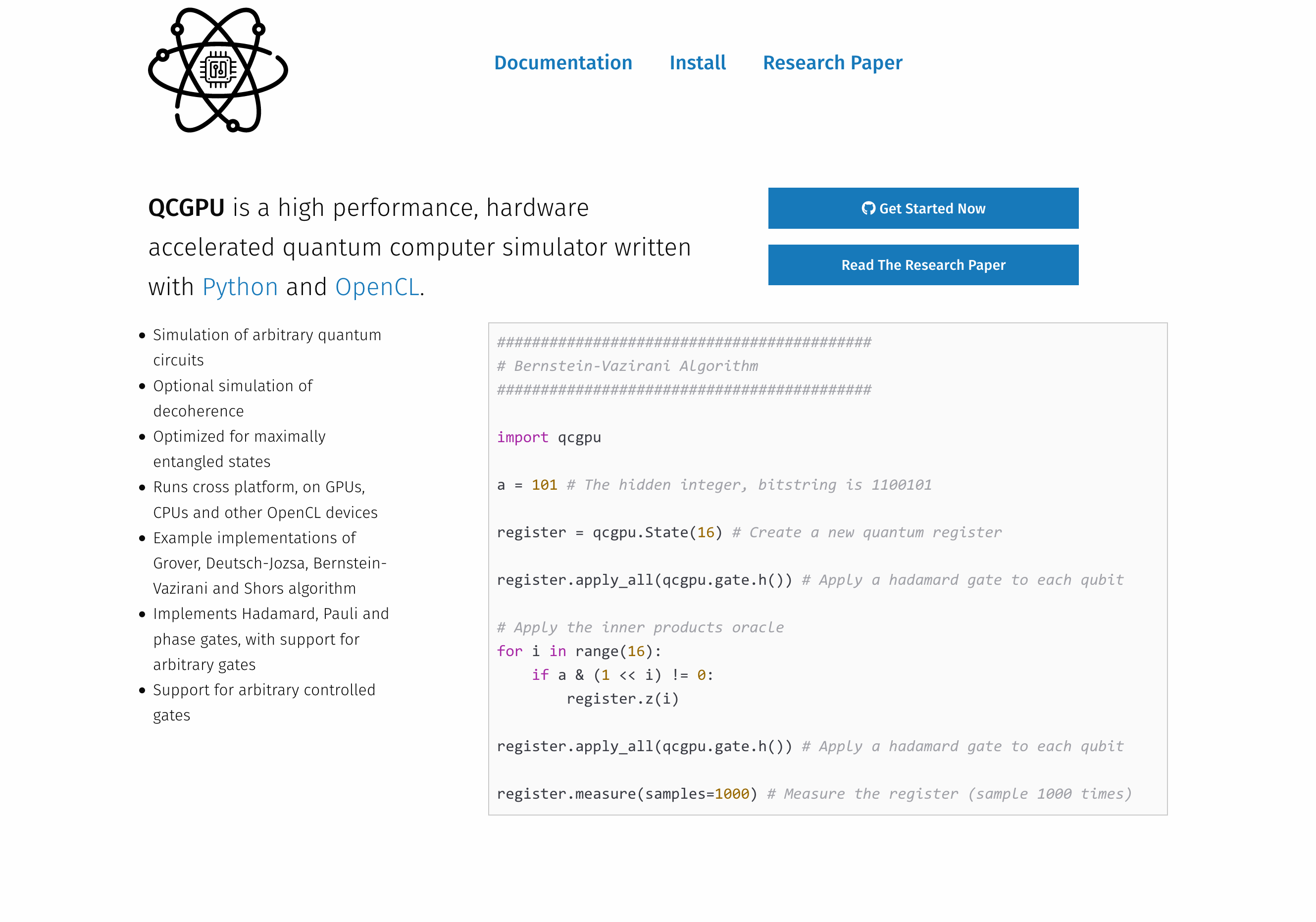}
	\caption{The website for the software library.}
	\label{fig:website}
\end{figure}

Because the software is open source, it is easily accessible (see figure \ref{fig:website}), thus enables it's use without having to get proprietary software, or pay some kind of subscription.
This means that any research done using the software (such as the simulation of algorithms) can be reproduced freely, and easily. It also lowers the
barrier to entry in regards to using the software.

The need to simulate quantum computers is likely one that will not go away, and will be essential to development of quantum devices.
The use of simulators is vital in the development of quantum algorithms also,
as it is the only way to have knowledge of what the internal state of the quantum computer would be like when running the algorithm.

Because of some of the features of this library (namely hardware acceleration), this library fits a wide range of use cases, especially those of
labs that already have this kind of hardware available (due to the popularity in fields such as machine learning, which also takes advantage of hardware acceleration).
The speedup offered by the hardware acceleration makes this library a valid choice for researchers in the theoretical and practical quantum computing field.

\subsection{Areas for Future Research}

There is many areas for future research in regards to this work.

Because quantum computers are described using linear algebra,
there exists a wide variety of ways (other than the state vector approach taken in this work)
to simulate quantum computers.
Some of these include using the Feynman path integral formulation of quantum mechanics \cite{boixo_simulation_2017, chen_classical_2018},
using tensor networks \cite{markov_simulating_2008}
and applying different simulations for circuits made up of certain types of quantum gates \cite{bravyi_simulation_2018}.
Graph-based approaches \cite{chen_64-qubit_2018} have also been shown as successful.
These techniques (while not covered in this work) will hopefully be included in the simulation software at a later date.

The simulator described in this report was able to simulate 28 qubits. To simulate more, a distributed approach would have to be taken.
These approaches are detailed in \cite{smelyanskiy_qhipster_2016, larose_distributed_2018}.

It is also planned to integrate the software with other quantum computing frameworks, to improve it's usefulness and
versatility.

% \subsection{Simulating Quantum Circuits} 

% \subsubsection{Challenges}
% \subsubsection{State-of-the-Art Solutions}
% \subsubsection{State Vector Simulators}

% \subsection{OpenCL} \label{opencl}

% \section{Conducting Simulations Using OpenCL}

% \subsection{Representation of State Vectors}

% \subsection{Representation of Gates}

% \subsection{Gate Application}

% \subsection{Measurement}

% \section{Comparison With Other Simulators}

% \subsection{Benchmark Implementation}

% \subsection{Results}

% \section{Conclusion}

% \subsection{Future Work}

% If you want a single column bibliography, comment below
\newpage
% \onecolumn
% \nocite{*
\bibliographystyle{abbrvnat}
\bibliography{references}
 
\onecolumn\newpage
\appendix

\section{Benchmarking Source Code}

The following is the source code used during the benchmarking of QCGPU against ProjectQ and Qiskit.

\begin{lstlisting}[language=Python]
  import click
  import time
  import random
  import statistics
  import csv
  import os.path
  import math
  
  from qiskit import QuantumRegister, QuantumCircuit
  from qiskit import execute, Aer
  
  from projectq import MainEngine
  from projectq.backends import Simulator
  import projectq.ops as ops
  
  import qcgpu
  
  def construct_circuit(num_qubits):
      q = QuantumRegister(num_qubits)
      circ = QuantumCircuit(q)
  
      # Quantum Fourier Transform
      for j in range(num_qubits):
          for k in range(j):
              circ.cu1(math.pi/float(2**(j-k)), q[j], q[k])
          circ.h(q[j])
  
      return circ
  
  
  # Benchmarking functions
  qiskit_backend = Aer.get_backend('statevector_simulator')
  eng = MainEngine(backend=Simulator(), engine_list=[])
  
  # Setup the OpenCL Device
  qcgpu.backend.create_context()
  
  def bench_qiskit(qc):
      start = time.time()
      job_sim = execute(qc, qiskit_backend)
      sim_result = job_sim.result()
      return time.time() - start
  
  def bench_qcgpu(num_qubits):
      start = time.time()
      state = qcgpu.State(num_qubits)
  
      for j in range(num_qubits):
          for k in range(j):
              state.cu1(j, k, math.pi/float(2**(j-k)))
          state.h(j)
  
      state.backend.queue.finish()
      return time.time() - start
  
  def bench_projectq(num_qubits):
      start = time.time()
  
      q = eng.allocate_qureg(num_qubits)
  
      for j in range(num_qubits):
          for k in range(j):
              ops.CRz(math.pi / float(2**(j-k))) | (q[j], q[k])
      ops.H | q[j]
      eng.flush()
  
      t = time.time() - start
      # measure to get rid of runtime error message
      for j in q:
          ops.Measure | j
  
      return t

  def benchmark(samples, qubits, out, single):
      functions = bench_qcgpu, bench_qiskit, bench_projectq
      times = {f.__name__: [] for f in functions}
      writer = create_csv(out)
  
      for n in range(0, qubits):
          # Construct the circuit
          qc = construct_circuit(n+1)
  
          # Run the benchmarks
          for i in range(samples):
              func = random.choice(functions)
              if func.__name__ != 'bench_qiskit':
                  t = func(n + 1)
              else:
                  t = func(qc)
              times[func.__name__].append(t)
  
  if __name__ == '__main__':
    benchmark()
\end{lstlisting}

\section{OpenCL Kernel Source Code}

\subsection{Gate Application}\label{adx:gate}

\begin{lstlisting}[language=C]

  /*
   * Returns the nth number where a given digit
   * is cleared in the binary representation of the number
   */
  static int nth_cleared(int n, int target)
  {
      int mask = (1 << target) - 1;
      int not_mask = ~mask;
  
      return (n & mask) | ((n & not_mask) << 1);
  }
  
  /*
   * Applies a single qubit gate to the register.
   * The gate matrix must be given in the form:
   *
   *  A B
   *  C D
   */
  __kernel void apply_gate(
      __global cfloat_t *amplitudes,
      int target,
      cfloat_t A,
      cfloat_t B,
      cfloat_t C,
      cfloat_t D)
  {
      int const global_id = get_global_id(0);
  
      int const zero_state = nth_cleared(global_id, target);
      int const one_state = zero_state | (1 << target);
  
      cfloat_t const zero_amp = amplitudes[zero_state];
      cfloat_t const one_amp = amplitudes[one_state];
  
      amplitudes[zero_state] = cfloat_add(cfloat_mul(A, zero_amp), cfloat_mul(B, one_amp));
      amplitudes[one_state] = cfloat_add(cfloat_mul(D, one_amp), cfloat_mul(C, zero_amp));
  }
  \end{lstlisting}

\subsection{Controlled Gate Application}\label{adx:control}

\begin{lstlisting}[language=C]
/*
* Applies a controlled single qubit gate to the register.
*/
__kernel void apply_controlled_gate(
    __global cfloat_t *amplitudes,
    int control,
    int target,
    cfloat_t A,
    cfloat_t B,
    cfloat_t C,
    cfloat_t D)
{
    int const global_id = get_global_id(0);
    int const zero_state = nth_cleared(global_id, target);
    int const one_state = zero_state | (1 << target); // Set the target bit

    int const control_val_zero = (((1 << control) & zero_state) > 0) ? 1 : 0;
    int const control_val_one = (((1 << control) & one_state) > 0) ? 1 : 0;

    cfloat_t const zero_amp = amplitudes[zero_state];
    cfloat_t const one_amp = amplitudes[one_state];

    if (control_val_zero == 1)
    {
        amplitudes[zero_state] = cfloat_add(cfloat_mul(A, zero_amp), cfloat_mul(B, one_amp));
    }

    if (control_val_one == 1)
    {
        amplitudes[one_state] = cfloat_add(cfloat_mul(D, one_amp), cfloat_mul(C, zero_amp));
    }
}
\end{lstlisting}

\subsection{Probability Calculation} \label{adx:measure}

\begin{lstlisting}[language=C]
  __kernel void calculate_probabilities(
      __global complex_f *const amplitudes,
      __global float *probabilities)
  {
      uint const state = get_global_id(0);
      complex_f amp = amplitudes[state];
  
      probabilities[state] = complex_abs(mul(amp, amp));
  }
\end{lstlisting}

\section{Example Implementation of the Bernstein-Vazirani Algorithm}

In this section, the Bernstein Vazirani algorithm is introduced,
along with it's implementation using the software developed in this project.

This algorithm was one of the first algorithms to show that quantum computers
could have a speedup over classical computers.
It shows the power of circuits that even have a low depth (not that many gates).

The implementation given here is without entanglement, 
and is based on a paper by Du et al. \cite{du_implementation_2000}.

\subsection{Introduction}

The Bernstein-Vazirani algorithm finds a hidden integer $a \in \{0, 1\}^n$
from an oracle $f_a$ that returns a bit $a \cdot x \equiv \sum_i a_i x_i \mod 2$ for an input $x$.

Implemented classically, the oracle returns $f_a (x) = ax \mod 2$.
The quantum oracle behaves analogously, but can be queried with a superposition.

To solve this problem classically, the hidden integer can be found by checking the
oracle with the inputs $x = 1,2,\dots,2^i,2^{n-1}$, where each
query reveals the $i$th bit of $a$ ($a_i$).
This is the optimal classical solution, and is $O(n)$. Using a quantum oracle and the
Bernstein-Vazirani algorithm, $a$ can be found with just one query to the oracle.

\subsection{Algorithm}

The Bernstein-Vazirani algorithm to find the hidden integer $a$ is very simple.
Start from the zero state $\ket{00 \dots 0}$, 
apply a Hadamard gate to each qubit,  
query the oracle,
apply another Hadamard gate to each qubit and measure the resulting state to find $a$.
This procedure is shown in algorithm \ref{alg:bernstein}.

\begin{algorithm}[h]
  \KwIn{A quantum oracle $U_{f_a}$ that returns a bit $a \cdot x \equiv \sum_i a_i x_i \mod 2$, for a hidden integer $a \in \{0, 1\}^n$ and input $x$}
  \KwOut{$a$: the hidden integer}
  \BlankLine
  $\ket{\psi} \leftarrow \ket{000 \dots 000}$\;
  $\ket{\psi} \leftarrow H^{\otimes n}$\;
  $\ket{\psi} \leftarrow U_{f_a}$\;
  $\ket{\psi} \leftarrow H^{\otimes n}$\;
  \Return{$a \leftarrow$ Measure $\ket{\psi}$}\;
%   \While{$\tau_r$ not found}
%   {
%     \For{$i \leftarrow 0$ \KwTo 3}
%     {
%       $\sigma_i \leftarrow$ get face opposite vertex $i$ in $\tau$\;
%       \If{Orient($\sigma_i, p$) $< 0$\nllabel{l:walk}} 
%       {
%         $\tau \leftarrow$ get neighbouring tetrahedron of $\tau$ incident to $\sigma_i$\;
%         break\;
%       }
%     }  
%     \If{$i=3$}
%     {
%       \tcp{all the faces of $\tau$ have been tested}
%       \Return{$\tau_r$ = $\tau$}
%     }
%   }
  \caption{B\textsc{ernstein-}V\textsc{azirani} ($f_a$)}
\label{alg:bernstein}
\end{algorithm}

The correctness of this algorithm can be shown too.
Consider the state $\ket{a}$, where measuring the state would result in
the binary string corresponding to the hidden integer $a$.
If a Hadamard gate is applied to each qubit in that state, the resulting state is

\begin{equation}
    \ket{a} \xrightarrow{H^{\otimes n}} \frac{1}{\sqrt{2^n}} \sum_{x\in \{0,1\}^n} (-1)^{a\cdot x} \ket{x}.
\end{equation}

Now consider the state $\ket{000\dots 0}$, the same state that the algorithm starts in.
Applying Hadamard gates gives

\begin{equation}
    \ket{000\dots 0} \xrightarrow{H^{\otimes n}} \frac{1}{\sqrt{2^n}} \sum_{x\in \{0,1\}^n} \ket{x}.
\end{equation}

These two states differ by a phase of $(-1)^{ax}$.

Now, the quantum oracle $f_a$ returns $1$ on input $x$ such that $a \cdot x \equiv 1 \mod 2$, and returns $0$ otherwise. 
This means we have the following transformation:

\begin{equation}
|x \rangle \left(|0\rangle - |1\rangle \right) \xrightarrow{f_a} | x \rangle \left(|0 \oplus f_a(x) \rangle - |1 \oplus f_a(x) \rangle \right) = (-1)^{a\cdot x} |x \rangle \left(|0\rangle - |1\rangle \right), 
\end{equation}
\noindent
where $\oplus$ is the XOR operation (outputs 1 only when the inputs differ) and $\ket{0} \equiv \ket{00\dots 0}$.
In the above equation, the $\ket{0} - \ket{1}$ state does not change, and can be ignored.
Thus, the oracle can create $(-1)^{ax}\ket{x}$ from the input $\ket{x}$.

With this, starting from the state $\ket{0}$,

\begin{align}
\ket{0}  & \xrightarrow{H^{\otimes n}} \frac{1}{\sqrt{2^n}} \sum_{x\in \{0,1 \}^n} \ket{x} \\
& \xrightarrow{f_a}   \frac{1}{\sqrt{2^n}} \sum_{x\in \{0,1 \}^n} (-1)^{ax} \ket{x} \\
& \xrightarrow{H^{\otimes n}}  \ket{a},
\end{align}
\noindent
as the Hadamard gates cancel.

\subsection{Inner Product Oracle}

The oracle used in this algorithm is the Inner product oracle. 
It transforms the state $\ket{x}$ into the state $(-1)^{ax} \ket{x}$.
The method of construction shown here requires no ancilla qubits (extra qubits not used in the final result) \cite{du_implementation_2000}. 
This is not the only method. Another approach is to use CNOT gates, but that does require ancilla qubits.

To construct the oracle, first note that

\begin{equation}
    (-1)^{a\cdot x} = (-1)^{a_1 x_1} \ldots (-1)^{a_ix_i} \ldots (-1)^{a_nx_n} = \prod_{i: a_i = 1} (-1)^{x_i}.
\end{equation}

It follows from this that the inner product oracle can be composed of single qubit gates,

\begin{equation}
O_{f_a} = O^1 \otimes O^2 \otimes \dots \otimes O^i \otimes \dots \otimes O^n,
\end{equation}
\noindent
where $O^i = (1 - a_i)I + a_i Z$. The gates $I$ and $Z$ are the identity gates and Pauli Z gates respectively,
and $a_i \in \{ 0, 1\}$.

\subsection{Implementation}

Now, an implementation of this algorithm using QCGPU
will be shown.

\begin{lstlisting}[language=Python]
import qcgpu
\end{lstlisting}

First, the number of qubits to use in the experiment can be set. 
Also, in order to construct the oracle, the hidden integer $a$ must be given.

\begin{lstlisting}[language=Python]
num_qubits = 14 # how many qubits to use
a = 101 # the hidden integer. bit-string is 1100101
\end{lstlisting}

Now the algorithm can be implemented

\begin{lstlisting}[language=Python]
# Create the quantum register
register = qcgpu.State(num_qubits)

# Apply Hadamard gates to each qubit
for i in range(num_qubits):
    register.h(i)

# Apply the inner-product oracle
for i in range(num_qubits):
    if (a & (1 << i)):
        register.z(i)
    # note: here should be an identity gate, 
    # but that doesn't modify the state

# Apply Hadamard gates to each qubit
for i in range(num_qubits):
    register.h(i)

# Measure the register
measurements = register.measure(samples = 1000)
\end{lstlisting}

\begin{figure}[h]
	\centering
	\includegraphics[width=0.35\linewidth]{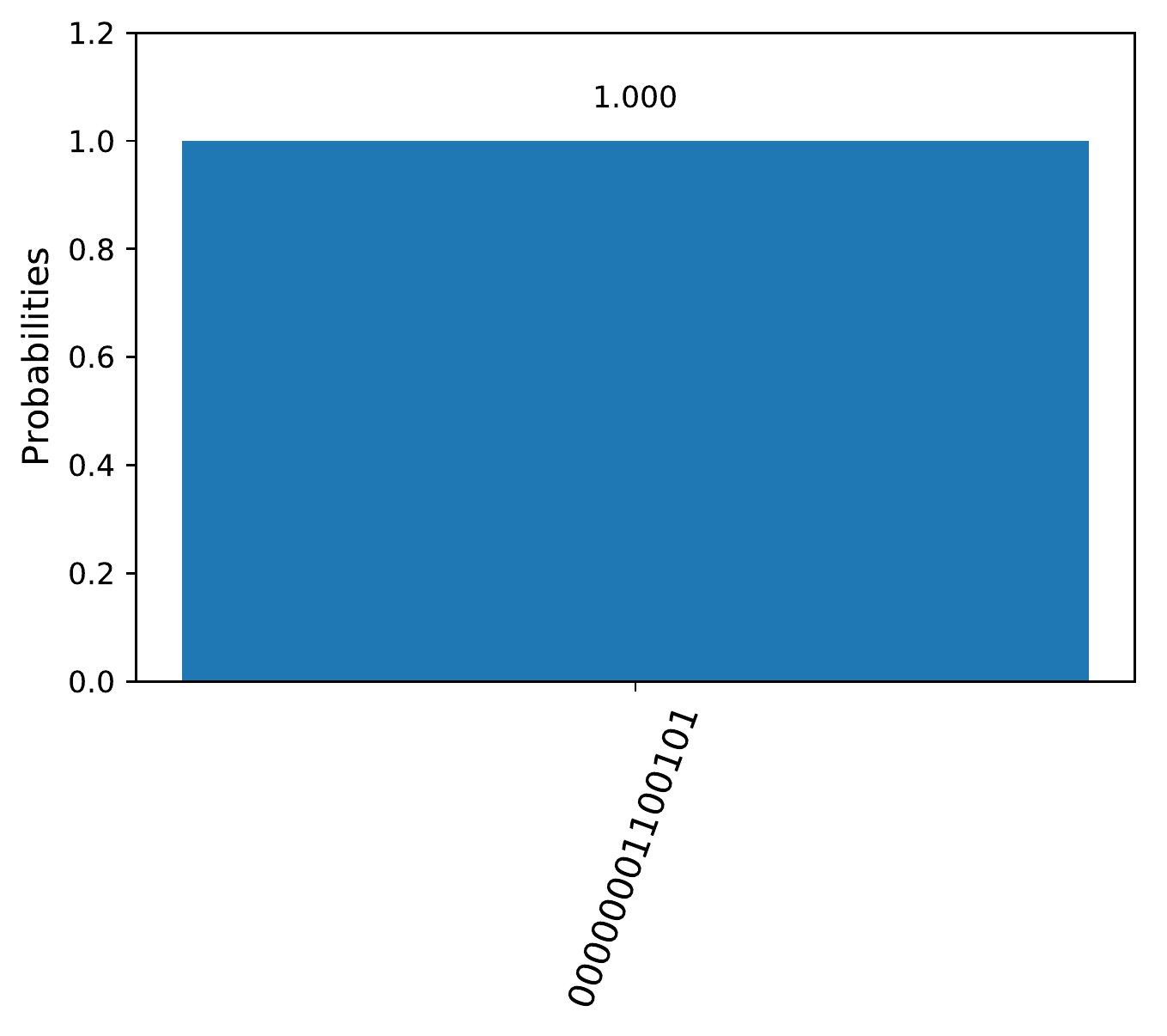}
  \caption{Bernstein-Vazirani Measurement Outcomes}
  \label{fig:bernstein_measurements}
\end{figure}

As can be seen from figure \ref{fig:bernstein_measurements}, 
the measurement outcome is the same as the bit-string of the hidden integer $a$.

\end{document}